\documentclass{jpsj3}
\usepackage{mathcomp}
\usepackage{color}
\usepackage{xcolor} 
\usepackage{bm}
\usepackage[version = 4]{mhchem}
\usepackage[resetlabels,labeled]{multibib}

\title{Microscopic Determination of the $c$-axis-Oriented Antiferromagnetic Structure in LaMnSi by $^{55}$Mn and $^{139}$La NMR}
\author{ Yusuke Sakai$^{1}$\thanks{sakai.yuusuke.88c@st.kyoto-u.ac.jp}, Fumiya Hori$^1$\thanks{Present address: Department of Physics, Tohoku University, Sendai 980-
8578, Japan}, Hiroki Matsumura$^{1}$, Shumpei Oguchi$^{1}$, Shunsaku Kitagawa$^{1}$\thanks{kitagawa.shunsaku.8u@kyoto-u.ac.jp}, Kenji Ishida$^1$, and Hiroshi Tanida$^{2}$}
\inst{
$^1$Department of Physics, Graduate School of Science, Kyoto University, Kyoto 606-8502, Japan \\
$^2$Liberal Arts and Sciences, Toyama Prefectural University, Imizu, Toyama 939-0398, Japan} 
\date{\today}
\abst{
We report a microscopic investigation of the magnetic structure and electronic properties of LaMnSi in its antiferromagnetic (AFM) state using nuclear magnetic resonance (NMR). Field-swept $^{55}$Mn- and $^{139}$La-NMR spectra, as well as zero-field $^{55}$Mn-NMR (ZFNMR) spectra, reveal that the Mn ordered moments are parallel to the tetragonal $c$ axis, consistent with the C-type AFM structure and the realization of an odd-parity multipole order. 
The internal field at the Mn site is determined to be 19.64 T at 4.2 K, corresponding to a hyperfine coupling constant of $A_{\rm hf} \sim 6.0~\mathrm{T/\mu_B}$. 
Nuclear spin-lattice relaxation rate $1/T_1$ exhibits a characteristic behavior of itinerant antiferromagnetism, showing metallic behavior at low temperatures and magnon-induced enhancement upon approaching the Néel temperature ($T_{\rm N} = 295$~K). 
These results show LaMnSi as an ideal compound to study $3d$ electron magnetism and odd-parity multipole order in the $RT$Si ($R$ = rare-earth, $T$ = transition metal) system, free of the complexities of $4f$ electrons.
}

\begin{document}
\maketitle

\section{Introduction}
\begin{figure}[tbp]
\begin{center}
\includegraphics[width=8cm]{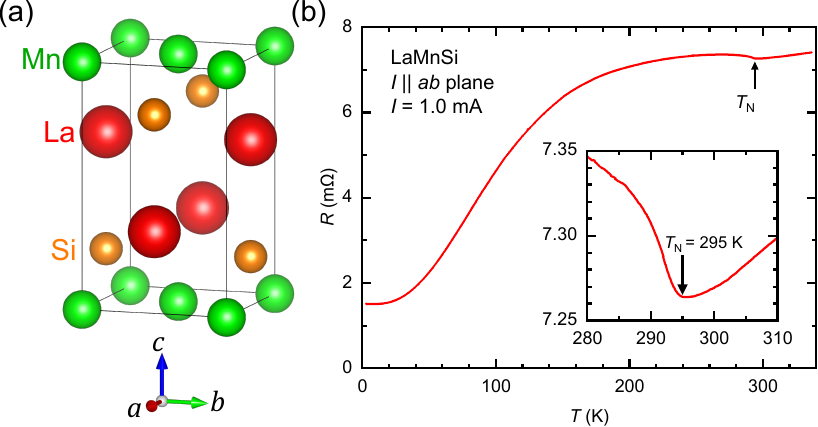}
\end{center}
\caption{(Color online) (a) Crystal structure of LaMnSi drawn by VESTA\cite{VESTA}.
(b) Temperature dependence of the in-plane electrical resistance of LaMnSi single crystal. The inset shows an enlarged view around $T_{\rm N}$.  }
\label{f1}
\end{figure}

Compounds of the $RT$Si family ($R$ = rare-earth, $T$ = transition metal) are known to exhibit remarkably diverse physical properties. Some of these compounds crystallize in the tetragonal CeFeSi-type structure (space group $P4/nmm$, No. 129)\cite{RTSi_structure_Bodak1970,GdMSi_structure_Kido1984,RMnSI_structure_Kido1986,RTXreview_Gupta2015} as shown in Fig. \ref{f1}(a). A characteristic feature of this structure is that space-inversion symmetry exists at the midpoint between atomic sites but is locally broken at the atomic sites. Namely, the structure possesses sublattice degrees of freedom that are exchanged under space-inversion operations, allowing for the emergence of odd-parity multipoles through staggered antiferromagnetic (AFM) order between sublattices on the rare-earth or transition-metal sites\cite{CeCoSi_odd-parity_multipole_yatsushiro2020}. Therefore, the $RT$Si family is of particular interest because its structural characteristics can host odd-parity multipole order in addition to the cooperative electronic states involving $3d$ and $4f$ electrons.

Many studies have focused on Ce-based systems, with CeCoSi being a notable example. CeCoSi exhibits a nonmagnetic phase transition at $T_0 \sim 12$~K, referred to as a “Hidden Order"\cite{CeCoSi_HOpressure_Lengyel2013,CeCoSi_HO_Tanida2019,CeCoSi_HO_Kimura2023}, followed by an AFM order of the Ce atoms at $T_{\rm N} \sim 9.4$~K\cite{CeCoSi_MO_and_HO_Chevalier2004,CeCoSi_magn_transition_Chevalier2006,CeCoSi_HO_Tanida2019,CeCoSi_magnetic_structure_Nikitin2020,CeCoSi_theory_Ishitobi}. The transition at $T_0$ has been discussed in terms of possible antiferroquadrupole order of Ce-$4f$ electrons\cite{CeCoSi_AFQ_Tanida2018,CeCoSi_HO_NMR_Manago2021,CeCoSi_odd-parity_multipole_yatsushiro2020,CeCoSi_NQRNMR_Yatsushiro2020,CeCoSi_HO_DFT_Yamada2024,CeCoSi_HO_theory_Hattori2025}. Moreover, X-ray diffraction experiments have reported a triclinic lattice distortion below $T_0$, suggesting a coupling between structural and electronic degrees of freedom\cite{CeCoSi_structural_transition_Matsumura2022,CeCoSi_phase_diagram_Hidaka2025}.

In contrast, CeMnSi, where the transition-metal site is replaced with Mn instead of Co, shows an AFM order of the Mn atoms at around 242 K\cite{CeMnSi_heavy-fermion_Tanida2023,RMnSi_neutron_Welter1994}. Recent bulk measurements, including specific heat and magnetic susceptibility, indicate the absence of magnetic order of Ce. Instead, a heavy-fermion state\cite{CeMnSi_heavy-fermion_Tanida2023} develops, which 
cannot be accounted for by the conventional Doniach's picture\cite{Doniach1977}.
In addition, it has been reported that a nontrivial phase is induced under magnetic fields, which makes the similarity to the hidden order phase in CeCoSi particularly intriguing.\cite{CeMnSi_Fieldtransition_Tanida2024}

Furthermore, in PrMnSi and NdMnSi, where Ce is replaced by other rare-earth elements, it has been reported that the Mn sublattice first develops AFM order, followed by AFM order of the rare-earth moments at lower temperatures\cite{RMnSi_neutron_Welter1994}. These examples illustrate that in the $RT$Si compounds, the electronic states of the transition-metal and rare-earth atoms mutually influence each other, giving rise to a rich variety of multiple ordered states.

Against this backdrop, LaMnSi is particularly compelling. Since La is nonmagnetic without $4f$ electrons, we can investigate the magnetism and electronic states that originate purely from Mn $3d$ electrons, without the complex effects arising from $4f$ electron interactions or quantum fluctuations. In other words, LaMnSi serves as an ideal reference compound for Ce- and other rare-earth-based $RT$Si compounds exhibiting cooperative $3d$–$4f$ phenomena, since its behavior can be understood purely in terms of $3d$ electrons.

A further motivation for studying LaMnSi is the potential realization of odd-parity multipoles associated with the Mn-AFM order. 
One intriguing aspect of multipole order is that various cross-correlated responses\cite{LightHelicity_Torque_Nukui2025,Piezomag_Naka2025} become active depending on the type of the ordered multipole. In the case of odd-time-reversal and odd-parity multipole order, phenomena such as current-induced strain\cite{CaMn2Bi2_MPE_Shiomi2019,EuMnBi2_MPE_Shiomi2020} and the magnetoelectric effect\cite{Cr2O3_ME_Astrov1960, MnTiO3_ME_Mufti2011, MEeffect_Co4Nb2O9_Yanagi2018, Toroidal_Hayami2014} have been theoretically proposed to become active\cite{BaMn2As2_Watanabe2017,PG_multipole_Watanabe2018,Multipole_calssification_Yatsushiro2021,CPG_multipole_Hayami2018,Multipole_Review_Hayami2024}.
In this context, LaMnSi, being free of $4f$ contributions, offers the special opportunity to probe the intrinsic effects of $3d$ multipole order.

Several previous studies on LaMnSi have been reported, but the magnetic structure remains under debate. Early neutron scattering experiments suggested the C-type AFM structure with the Néel vector tilted by 45° from the $c$ axis\cite{RMnSi_neutron_Welter1994}, whereas the anisotropy of bulk magnetization measurements points to the C-type AFM structure with the Néel vector parallel to the $c$ axis\cite{LaMnSi_Tanida2022}, making the magnetic structure controversial. In CeMnSi, by contrast, the Néel vector in the Mn-AFM state is reported to be perpendicular to the $c$ axis\cite{CeCoSi_magnetic_structure_Nikitin2020}, differing from either of the proposed magnetic structures in LaMnSi. Therefore, determining the magnetic structure of LaMnSi is crucial for understanding how the choice of rare-earth element influences the Mn-AFM order in $R$MnSi.

In this paper, we present a microscopic investigation of the magnetic structure and electronic properties of LaMnSi in its AFM state using nuclear magnetic resonance (NMR) measurements. Our results reveal that in the Mn-AFM state, the magnetic ordered moments are aligned parallel to the $c$ axis, confirming the realization of an odd-parity multipole ordered state. Furthermore, the magnetic fluctuations exhibit conventional metallic behavior at low temperatures, followed by a divergent increase as the system approaches the Néel temperature, characteristic of itinerant magnetism. 
These findings not only provide an important foundation for discussing the contribution of rare-earth $4f$ electrons in the $RT$Si system, but also establish LaMnSi as an odd-parity multipole ordered system.
In particular, its metallic nature makes it a promising platform for exploring current-driven multipole phenomena.

\section{Experimental}
Single-crystal LaMnSi samples were grown by the self-flux method. The crystal size was $3.5 \times 2.3 \times 0.2$ mm$^3$, with a thin plate-like shape along the $ab$ plane. NMR measurements were performed using the spin-echo technique. For NMR under an external magnetic field, the field was applied along the $c$ axis.
$^{55}$Mn-NMR spectra (nuclear spin 
$I$ = 5/2, gyromagnetic ratio $^{55}\gamma/2\pi$ = 10.554 MHz/T, natural abundance 100 \%) and 
$^{139}$La-NMR spectra (nuclear spin $I$ = 7/2, gyromagnetic ratio $^{139}\gamma/2\pi$ = 6.0142 MHz/T, natural abundance 99.91 \%) were obtained as a function of magnetic field at fixed frequencies of 65.15, 78.18, and 84.00 MHz. For magnetic field calibration, $^{63}$Cu ($^{63}\gamma /2\pi = 11.285$ MHz/T) and $^{65}$Cu ($^{65}\gamma /2\pi = 12.089$ MHz/T) NMR signals with the Knight shift $K_{\rm Cu}=0.238(5)\%$ from the NMR coil were used\cite{GCCarter1976}. Subsequently, the sample broke into three pieces during the experiment, and zero-field NMR (ZFNMR) measurements were performed by stacking the three pieces to obtain high-intensity $^{55}$Mn-ZFNMR spectra.
For electrical resistance measurements, the largest piece ($2.5 \times 1.8 \times 0.2$ mm$^3$) was used. Electrical contacts were made by spot welding, and the resistance was measured using the four-terminal method. The single crystal sample exhibited an AFM order at $T_{\rm N} = 295$~K, determined from the resistance measurements shown in Fig. \ref{f1}(b).
The nuclear spin-lattice relaxation rate $1/T_{\rm 1}$ of $^{55}$Mn and $^{139}$La was determined by fitting the time evolution of the spin-echo intensity after the saturation of the nuclear magnetization to the theoretical functions for $I$ = 5/2 and $I$ = 7/2 under zero field and at 10.815 T, respectively. 
All NMR measurements were performed at temperatures below $T_{\rm N}$, corresponding to the AFM ordered phase.

\begin{figure*}[t]
\begin{center}
\includegraphics[width=15cm]{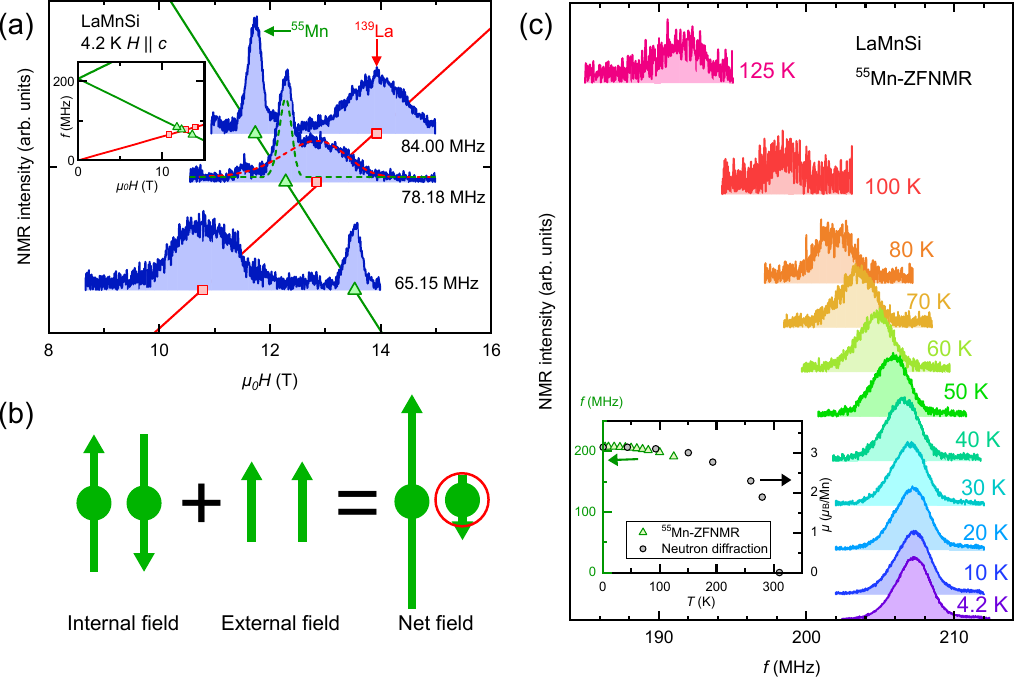}
\end{center}
\caption{(Color online) (a) Field-swept NMR spectra obtained at 4.2 K for three different resonance frequencies.
The peak position of the $^{139}$La-NMR signal is indicated by red squares, and that of the $^{55}$Mn-NMR signal is indicated by yellow-green triangles.
Both were observed as broad single peaks. The vertical offset between spectra is scaled according to the frequency interval. The inset shows the extrapolation of the linear fit to zero magnetic field, along with the expected resonance frequency of the $^{55}$Mn-ZFNMR signal originating from the other Mn sublattice with an oppositely oriented internal field. (b) Schematic illustration explaining the relation between the resonance field and resonance frequency in $^{55}$Mn-NMR. The Mn site highlighted in red corresponds to the $^{55}$Mn-NMR signal observed in (a). (c) Temperature dependence of the $^{55}$Mn-ZFNMR spectrum from 4.2 K to 125 K. The inset shows the temperature dependence of the resonance frequency of $^{55}$Mn-ZFNMR (light-green triangles) and the ordered Mn moment (black circles)\cite{RMnSi_neutron_Welter1994}.}
\label{f2}
\end{figure*}

\section{Results}
To determine the magnetic structure of LaMnSi, the relation between the NMR resonance frequency and the resonance magnetic field was investigated. Figure \ref{f2}(a) shows the field-swept spectra obtained at three different frequencies at 4.2 K. At each frequency, NMR signals from $^{55}$Mn and $^{139}$La were observed.
Since both $^{55}$Mn and $^{139}$La have nuclear spin $I \ge 1$, their nuclei should experience not only the Zeeman interaction between the magnetic field and the nuclear spin, but also a finite nuclear quadrupole interaction with the electric field gradient. In general, such conditions are expected to produce NMR spectra with multiple peaks. 
However, in the present measurements, the $^{55}$Mn and $^{139}$La signals both appeared as single broad peaks. This suggests that inhomogeneity in the local environment within the crystal smears out the multiple peaks, resulting in a single broadened resonance peak.
Thus, an improvement in sample quality could enable the observation of five and seven resonance peaks split by the quadrupole interaction.
The inset of Fig. \ref{f2}(a) shows linear fits to the relation between the resonance field and resonance frequency for $^{139}$La and $^{55}$Mn.
Focusing first on the $^{139}$La signal, the linear fit yields an intercept close to zero, and the slope is found to be 6.0(3) MHz/T, which is nearly equal to the nuclear gyromagnetic ratio of $^{139}$La (6.0142 MHz/T).
Similarly, a linear fit to the $^{55}$Mn signal gives a slope of -10.47(3) MHz/T, whose absolute value is nearly equal to the nuclear gyromagnetic ratio of $^{55}$Mn (10.554 MHz/T). Moreover, the intercept value of 206.8(3) MHz indicates that the resonance frequency of the $^{55}$Mn-ZFNMR signal is approximately 207 MHz at 4.2 K.

The relation between the resonance field and the resonance frequency for $^{55}$Mn can be explained by the scenario shown in Fig. \ref{f2}(b). Since Mn is a magnetic atom, a large internal magnetic field exists in the AFM state. For simplicity, we assume here that the internal field forms a staggered structure parallel to the $c$ axis. 
Applying an external field along the $c$ axis increases the net field at Mn sites where the external and internal fields are parallel. Consequently, the resonance frequency shifts to higher values with increasing external field. Conversely, at the Mn sites where the external and internal fields are antiparallel, the resonance frequency decreases with respect to the external field. Thus, the $^{55}$Mn-NMR signal observed in the present study originates from the Mn sites where the internal and external fields are antiparallel. The actual direction of the internal field will be discussed later.

Figure \ref{f2}(c) shows the temperature dependence of the $^{55}$Mn-ZFNMR spectra. 
The $^{55}$Mn-ZFNMR signal was observed at the resonance frequency expected from the $^{55}$Mn-NMR spectra measured under an external magnetic field.
With increasing temperature, the resonance frequency shifts toward lower frequencies. The peaks exhibit a slightly asymmetric shape with a tail extending to the low-frequency side. In addition, the spectra become noisy at higher temperatures due to the reduced signal intensity. The resonance frequency in the $^{55}$Mn-ZFNMR spectrum reflects the magnitude of the internal field. In other words, the temperature dependence of the resonance frequency can be regarded as the temperature dependence of the order parameter, showing that the Néel vector decreases in magnitude as temperature increases. As shown in the inset of Fig. \ref{f2}(c), the temperature dependence of the resonance frequency is in good agreement with that of the ordered moment determined by neutron scattering\cite{RMnSi_neutron_Welter1994}. A fit of the resonance frequency to the expression
\begin{equation}
 f=f_0(1-cT^\alpha)
\end{equation}
gave $\alpha=2.47(4)$.
According to spin-wave theory, $\alpha=3/2$ for a three-dimensional ferromagnet\cite{FM_alpha_Bloch1930} and 
$\alpha=2$ for a three-dimensional antiferromagnet\cite{AFM_alpha_Kubo1952}. From this viewpoint, although the present value is somewhat larger, it is plausible.

Using the resonance frequency of 207.3 MHz at 4.2 K and the nuclear gyromagnetic ratio $^{55}\gamma/2\pi$ = 10.554 MHz/T, the internal magnetic field at the Mn site was determined to be 19.64 T at 4.2 K. 
From this, if we assume the Mn magnetic ordered moment to be 3.27(4)~$\mu_{\rm B}$ at 1.4~K as obtained from neutron diffraction\cite{RMnSi_neutron_Welter1994}, the hyperfine coupling constant is calculated as $A_{\rm hf} \sim 6.0~\mathrm{T/\mu_B}$. Hyperfine coupling constants have also been investigated for BaMn$_2Pn_2$ ($Pn$ = As, Sb, Bi), which exhibit $c$-axis-collinear Mn-AFM order, and values of 5.84, 6.60, and 5.51 $\mathrm{T/\mu_B}$ have been reported, respectively\cite{BaMn2Pn2_NMR_jansa2021}. Therefore, the value of $A_{\rm hf} \sim 6.0~\mathrm{T/\mu_B}$ for LaMnSi is reasonable.

\begin{figure}[tbp]
\begin{center}
\includegraphics[width=8cm]{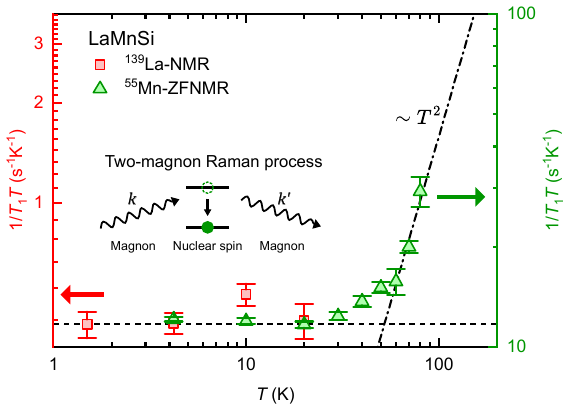}
\end{center}
\caption{(Color online) $T$ dependence of the nuclear spin-lattice relaxation rate $1/T_{\rm 1}T$ obtained from $^{55}$Mn-ZFNMR and $^{139}$La-NMR measurements.
The dashed and dash-dotted lines represent a temperature-independent behavior following the Korringa law and a $T^2$-dependent behavior, respectively. The inset illustrates the two-magnon Raman process, which accounts for the $T^2$-dependent behavior.}
\label{f3}
\end{figure}

The characteristics of itinerant antiferromagnetism can also be identified from the behavior of spin fluctuations. Figure \ref{f3} shows the temperature dependence of the nuclear spin-lattice relaxation rate divided by temperature, $1/T_{\rm 1}T$, for $^{55}$Mn and $^{139}$La. 
For $^{139}$La, measurements were performed at 65.15 MHz and 10.815 T over the temperature range from 1.5 K to 20 K. As a result, $1/T_{\rm 1}T$ follows the Korringa relation and exhibits almost no temperature dependence up to 20 K, which is typical of a conventional metal. 
For $^{55}$Mn, measurements were carried out in zero field using the resonance frequency at each temperature in the range of 4.2 K to 80 K. Up to 30 K, $1/T_{\rm 1}T$ remains constant as in the case of $^{139}$La. Notably, the constant value of $1/T_{\rm 1}T$ for $^{55}$Mn is more than twenty times larger than that for $^{139}$La. In relaxation processes dominated by magnetic fluctuations, the constant value of $1/T_{\rm 1}T$ is proportional to $A_{\rm hf}^2\gamma^2D(\epsilon_{\rm F})^2$, where $D(\epsilon_{\rm F})$ is the electronic density of states at the Fermi level. Taking into account the ratio of the nuclear gyromagnetic factors, $^{55}\gamma/^{139}\gamma$~=1.75(1), we obtain $^{55}A_{\rm hf}/^{139}A_{\rm hf} \sim 3$, indicating that the hyperfine coupling constant is larger for the magnetic Mn site.

In contrast, above 40 K, $1/T_{\rm 1}T$ shows a deviation from constancy and increases with increasing temperature. This behavior corresponds to the enhancement of magnon excitations as the system approaches the magnetic transition point. 
In particular, the $T^2$-dependence of $1/T_{\rm 1}T$ observed above 60 K suggests that the two-magnon Raman process predominantly drives the nuclear spin relaxation in this temperature range\cite{Magnon_T1_Beeman1968}.

\begin{figure}[tbp]
\begin{center}
\includegraphics[width=8cm]{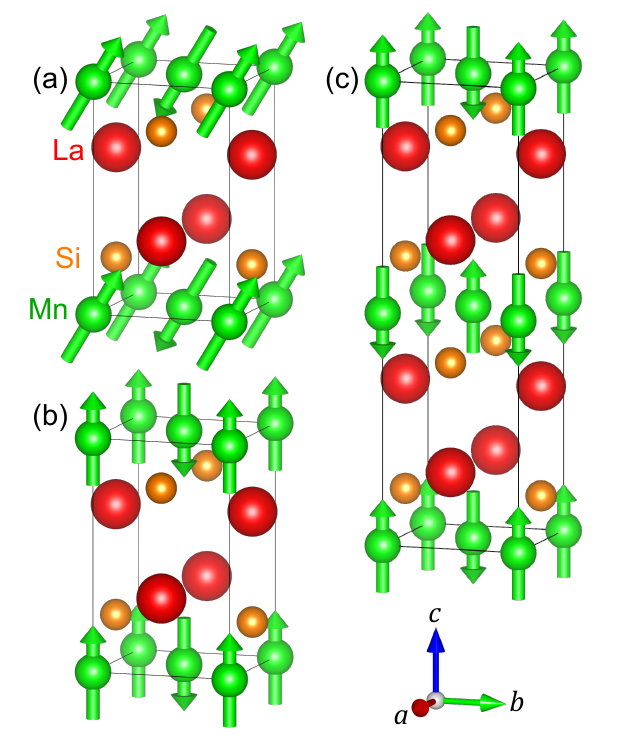}
\end{center}
\caption{(Color online) Three types of possible magnetic structures of LaMnSi drawn by VESTA\cite{VESTA}. (a) Magnetic structure suggested by neutron diffraction experiments on polycrystalline samples, in which the ordered moments are tilted by 45° from the $c$ axis.\cite{RMnSi_neutron_Welter1994} (b) Magnetic structure suggested by magnetic susceptibility measurements on single-crystal sample, in which the ordered moments are parallel to the $c$ axis.\cite{LaMnSi_Tanida2022} This structure is consistent with the present NMR results. (c) G-type AFM structure compatible with both the magnetic susceptibility\cite{LaMnSi_Tanida2022} and NMR results. 
 }
\label{f4}
\end{figure}

\section{Discussion}
Here, we discuss the magnetic structure of the AFM state in LaMnSi. At present, two different magnetic structures have been proposed in previous studies, which are shown in Fig. \ref{f4}(a) and (b), respectively\cite{RMnSi_neutron_Welter1994, LaMnSi_Tanida2022}: 

\begin{itemize}
    \item[(a)]C-type AFM structure with the Mn ordered moment tilted by 45° from the $c$ axis, suggested by neutron scattering\cite{RMnSi_neutron_Welter1994} 
  \item[(b)]C-type AFM structure with the Mn ordered moment aligned along the $c$ axis, inferred from the anisotropy of bulk magnetic susceptibility\cite{LaMnSi_Tanida2022}
\end{itemize}

Based on these scenarios, we now examine the magnetic structure consistent with our NMR results.
From Fig. \ref{f2}(a), it is first noted that the resonance frequency of $^{139}$La is determined by $^{139}\gamma H$, and the Knight shift is nearly zero. This indicates no internal field at the La site. 
Next, we analyzed the relation between the resonance frequency $f$ and resonance field of $^{55}$Mn $H$ by fitting the data. The fitting parameters were the magnitude of the internal field $H_{\rm int}$ and the angle $\theta$ between the internal and external fields, using the expression
\begin{equation}
  \begin{split}
    f &= \gamma |\boldsymbol{H}+\boldsymbol{H}_{\rm int}|\\
    &= \gamma \sqrt{H^2+H_{\rm int}^2+2HH_{\rm int}\cos{\theta}}
  \end{split}
\end{equation}
In magnetic structure (a), Mn sites with $\theta=45\tcdegree$ and $135\tcdegree$ exist, whereas in (b), Mn sites with $\theta=0\tcdegree$ and $180\tcdegree$ exist. From the fitting, we obtained $\theta=177.4(4) \tcdegree$. Considering that a few degrees of error are expected due to sample alignment, the $^{55}$Mn-NMR signal observed under the $c$ axis magnetic field is likely to originate from the Mn site with $\theta=180\tcdegree$. In other words, this strongly suggests magnetic structure (b).
Furthermore, unlike in case (a), structure (b) is also consistent with the absence of an internal field at the La site. The absence of an internal magnetic field at the La site in (b) can be understood from the fact that the La site in magnetic structure (b) is symmetric under the combined operation of a fourfold rotation around the $c$ axis and time reversal.

Although the NMR results support magnetic structure (b), which differs from the neutron scattering results \cite{RMnSi_neutron_Welter1994}, there remains another possible structure that can account for both the NMR data and the anisotropy of the bulk magnetic susceptibility \cite{LaMnSi_Tanida2022}, which is shown in Fig. \ref{f4}(c): 

\begin{itemize}
    \item[(c)]G-type AFM structure with the Mn ordered moment aligned along the $c$ axis
\end{itemize}

In this case, the propagation vector of the magnetic order is 
$\boldsymbol{q}=(0,0,0.5)$, in contrast to 
$\boldsymbol{q}=\boldsymbol{0}$ in cases (a) and (b). However, this scenario is unlikely. Comprehensive neutron scattering studies on $R$MnSi ($R$ = La, Ce, Pr, Nd)\cite{RMnSi_neutron_Welter1994} have shown that for $R$ = Pr and Nd the Mn-AFM order is characterized by $\boldsymbol{q}=(0,0,0.5)$. Since the diffraction patterns in such systems differ clearly from that of $R$ = La, the assignment of the $\boldsymbol{q}=\boldsymbol{0}$ order for LaMnSi can be considered reliable.
Although the $\boldsymbol{q}=\boldsymbol{0}$ order is strongly suggested, neutron scattering experiments on single crystals would be desirable to confirm it more conclusively.

Notably, in the case of magnetic structure (c), the magnetic order does not reduce the symmetry in terms of the magnetic point group, remaining in $4/mmm1'$. By contrast, in case (b), the symmetry is lowered to $4'/m'm'm$, which allows several cross-correlated responses to become active\cite{Multipole_calssification_Yatsushiro2021}. For example, from the viewpoint of symmetry, a magnetoelectric effect of the form $M_a = \alpha E_a$, where the magnetic moment $M_a$ responds linearly to the electric field $E_a$ along the $a$ axis, is allowed. Therefore, observation of such cross-correlated responses would provide further confirmation of the C-type AFM structure. It should also be noted that in CeMnSi, a $\boldsymbol{q}=\boldsymbol{0}$ Mn-AFM order with spins lying in the $ab$ plane has been proposed\cite{CeCoSi_magnetic_structure_Nikitin2020}.
That is, the present results highlight that the type of rare-earth element likely plays a crucial role in modifying the spin state of Mn $3d$ electrons.
The origin of this difference in the Mn spin orientation requires further investigation.

\section{Conclusion}
In conclusion, we have performed the $^{55}$Mn and $^{139}$La NMR measurements on LaMnSi
to clarify its AFM structure and electronic behavior. The Mn moments form the C-type AFM order with spins aligned along the $c$ axis, realizing an odd-parity multipole ordered state. Spin-lattice relaxation measurements indicate the character of Mn $3d$ electrons, showing conventional metallic behavior at low temperatures and enhanced spin fluctuations approaching $T_{\rm N}$. LaMnSi thus provides a clean platform to investigate intrinsic $3d$ magnetism and the emergence of odd-parity multipoles in metallic systems, offering a benchmark for understanding more complex Ce- and other rare-earth-based $RT$Si compounds.

\section*{Acknowledgments}
This work was supported by Grants-in-Aid for Scientific Research (KAKENHI Grant No. JP20KK0061, No. JP20H00130, No. JP21K18600, No. JP22H04933, No. JP22H01168, No. JP23H01124, No. JP23K22439, No. JP23K25821, No. JP25H00609 and No. JP25K07209) from the Japan Society for the Promotion of Science, by research support funding from the Kyoto University Foundation, by ISHIZUE 2024 of Kyoto University Research Development Program, by Murata Science and Education Foundation, and by the JGC-S Scholarship Foundation.
Liquid helium is supplied by the Low Temperature and Materials Sciences Division, Agency for Health, Safety and Environment, Kyoto University.

\bibliographystyle{jpsj_QM}

\bibliography{LaMnSi_ref}

\end{document}